\documentclass[aps,pra,twocolumn,showpacs,preprintnumbers,amsmath,footinbib,10pt]{revtex4-1}
\pdfoutput=1

\usepackage{graphicx}
\usepackage{dcolumn}
\usepackage{bm}
\usepackage[utf8]{inputenc}
\usepackage[T1]{fontenc}
\usepackage{mathptmx}
\usepackage{epsfig}
\usepackage{epstopdf}
\usepackage{amsmath}
\usepackage{txfonts}
\usepackage[sort&compress]{natbib}
\usepackage{rotating}
\usepackage{graphics}
\usepackage{array,multirow}
\usepackage[english]{babel}
\usepackage{attachfile}
\usepackage{pgfplots}
\usepackage[export]{adjustbox}
\usepackage{braket}
\usepackage{starfont}

\pgfplotsset{compat=1.15}
\hypersetup{urlcolor=purple, citecolor=purple, linkcolor=purple, colorlinks=true}

\def\be{\begin{equation}}
\def\ee{\end{equation}}
\def\bea{\begin{eqnarray}}
\def\eea{\end{eqnarray}}

\newcommand{\santi}[1]{{\color{black} #1}}

\begin{document}


\title{Magnetic pulses enable multidimensional optical spectroscopy of dark states}

\author{Santiago Oviedo-Casado${}^{1,2,{\text{\Chiron}}}$,Franti\v{s}ek \v{S}anda${}^{3,4,{\text{\Hidalgo}}}$,J{\"u}rgen Hauer${}^{4,5,{\text{\Hygiea}}}$,Javier Prior${}^{2,6,{\text{\Psyche}}}$}
\affiliation{${}^{1}$Racah Institute of Physics, The Hebrew University of Jerusalem, Jerusalem, 91904, Givat Ram, Israel\\
${}^{2}$Departamento de F{\'i}sica Aplicada, Universidad Polit{\'e}cnica de Cartagena, Cartagena 30202 Spain\\
${}^{3}$Institute of Physics, Faculty of Mathematics and Physics, Charles University, Ke Karlovu 5, Prague, 121 16 Czech Republic\\
${}^{4}$Fakult{\"a}t f{\"u}r Chemie, TU M{\"u}nchen, Oettingenstra{\ss}e 67, 80538 Munich, Germany\\
${}^{5}$Photonics Institute, TU Wien, Gu{\ss}hausstra{\ss}e 27-29, 1040 Vienna, Austria\\
${}^{6}$Instituto Carlos I de F{\'i}sica Te{\'o}rica y Computacional, Universidad de Granada, Granada 18071, Spain}%
\email{${}^{{\text{\Chiron}}}$oviedo.cs@mail.huji.ac.il\\
${}^{{\text{\Hidalgo}}}$sanda@karlov.mff.cuni.cz\\
${}^{{\text{\Hygiea}}}$juergen.hauer@tum.de\\
${}^{{\text{\Psyche}}}$javier.prior@upct.es}

\begin{abstract}
The study and manipulation of low dipole moment quantum states has been challenging due to their inaccessibility by conventional spectroscopic techniques. Controlling the spin in such 
states requires unfeasible strong magnetic fields to overcome typical decoherence rates. However, the advent of terahertz technology and its application to magnetic pulses opens up a new 
scenario. In this article, we focus on an electron-hole pair model to demonstrate that it is possible to control the precession of the spins and to modify the transition rates to different spin states. Enhancing transitions from a bright state to a dark state with different spin means that the latter can be revealed by ordinary spectroscopy. We propose a modification of the standard two-dimensional 
spectroscopic scheme in which a three pulse sequence is encased in a magnetic pulse. Its role is to drive transitions between a bright and a dark 
spin state, making the latter susceptible to spectroscopic investigation.
\end{abstract}

\maketitle

\section{Introduction}

Quantum states with small (zero) transition dipole moment with respect to the ground state, referred to as dark states, are notoriously difficult to investigate by linear spectroscopic techniques. The most widespread approach is detection via their low emission yields \cite{Cook1985,Yip1998}. Alternative methods are fluorescence blinking, transient absorption, and comparisons with modified samples suppressing the presumed dark state \cite{Ferretti2016,Krecik2015,Berkeland2002}. The fact that whole classes of electronic transitions are dipole forbidden lends fundamental relevance to their study. Dark states are often very stable and keep phase relation with other states, leading to small decoherence rates. Additionally, dark states are essential for understanding energy and charge transport phenomena in fields ranging from quantum optics to solid state physics. 

Examples are singlet - triplet spin structures, appearing from interactions of two spin 1/2 particles; triplet states are dark for dipole transitions from the singlet ground state. \santi{Here we focus on the class of states which are dark due to spin selection rules (henceforth spin dark states), in contrast to dark states which are dark owing to parity selection rules. Spin dark states such as the triplets investigated here have been demonstrated to be relevant for charge separation in artificial light harvesting, where they might hold the key to enhanced photocurrent generation \cite{Rao2013a,Chang2015a}}. Furthermore, they are responsible for efficient transport within quantum wells and two-dimensional materials, or for optical control in solid-state and semiconductor systems \cite{Snoke2002,Ye2014,Yale2013}. The control of spin dark states, in combination with advancing me\-thods of spintronics, holds the potential to all-optical control of such quantum systems. This line of research offers intriguing possibilities in quantum technological applications, such as quantum computing or sensing.

One possibility to overcome the dipole-forbidden character of dark states is to employ field-matter interactions of higher order, i.e. multipoles. However, this approach is hampered by the high field strengths necessary, making material damage threshold a limiting factor.

Instead of utilizing higher electric multipoles of dark states, we focus on their magnetic dipole moments. For triplet states, which are the subject of our research, the origin of magnetic dipole moments are their characteristic unpaired spins. Conventional techniques for the study of triplet states such as electron paramagnetic resonance (EPR) and multidimensional variants thereof are highly sensitive and insightful in terms of electronic structure determination and reactivity \cite{Chechik}. With respect to time-resolution, EPR-experiments are typically limited to the microsecond regime due to restrictions in microwave pulse technology. In the femtosecond regime, time resolved Two-Dimensional Electronic Spectroscopy (2DES) has emerged as the most comprehensive technique for the study of dynamics, due to its versatility and by its main characteristic of resolving a non-linear signal in excitation and emission frequencies \cite{Jonas2003}. Dark states occurring during relaxation can be detected here by their (potential) transitions to higher lying states, i.e. as excited state absorption. It has to be noted however that such ESA-features are often broad, featureless, convoluted, and therefore hard to analyze \cite{Polivka2004,Read2009,Perlik2015}. \santi{For some materials, singlet fission leads to allowed formation of triplet states which can be detected via 2DES \cite{Bakulin2016}.} In this article, we will propose a modification of 2DES allo\-wing to observe spin-dependent dark states. We argue that, apart from the linear spectroscopic signal and ESA-signatures, spin properties of quantum states can be exploited to make dark states responsive to direct spectroscopic probing. 

\begin{figure}
\centering
\includegraphics[width=\columnwidth]{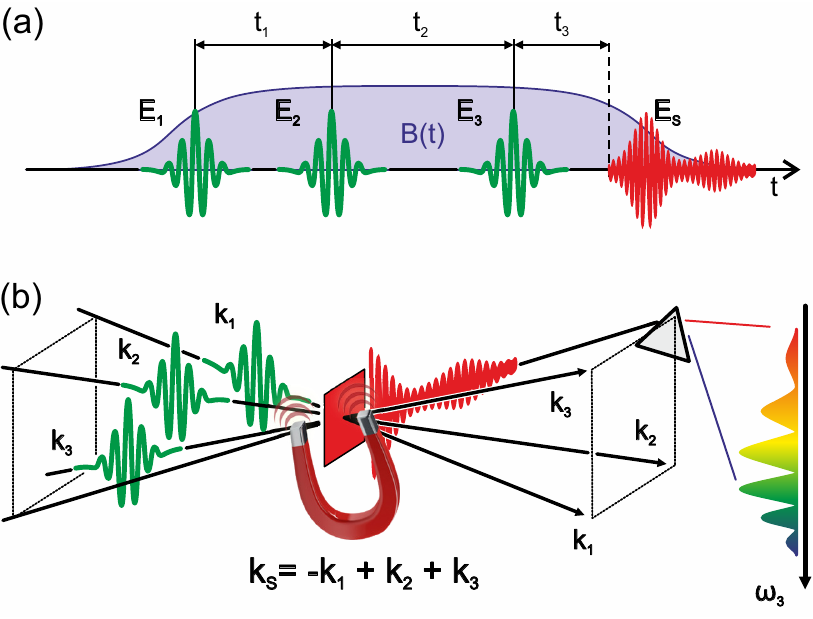}
\caption{Schematic representation of the proposed magnetic field enhanced two-dimensional spectroscopic experiment. In (a) the three laser pulse sequence is shown. The first pulse excites a bright state in the system, which evolves freely for a variable time $t_1$ (coherence time), during which the magnetic pulse (which acts all through the laser pulses sequence) induces coherent transitions from the excited bright state to a dark state. The second pulse stabilizes the population in the dark state while the third produces a rephasing signal. (b) displays the fully non-collinear phase matching geometry of the laser pulses which in our design are immersed in a magnetic pulse (violet filled curve).}
\label{Fig1}
\end{figure}

It was recently demonstrated that a static magnetic field can manipulate charge transfer states, owing to the non-trivial spin and small Coulomb binding properties \cite{Oviedo2017}. Moreover, these 
states can be inferred through the behavior of the resultant photocurrent when affected by magnetic fields of varying intensity \cite{Oviedo2018}. In particular, it was shown that a magnetic field 
can be used to induce dipole moment redistribution between a bright singlet charge transfer state and a dark triplet charge transfer state, making the latter bright. This result suggests that the same principle could be applied to any 
such dark state that is spin-connected to a bright state. Recently, static magnetic fields as high as 25 Tesla have been employed to study energy transport and charge transfer in biomolecules \cite{Scholes2018}. Despite their immense field strength, such static fields are not capable of modifying inter-system crossing (i.e. singlet-triplet transitions) in a time-scale relevant for 2DES, as spin precession depends on the coupling to the magnetic field, which is controlled by the Bohr magneton, and limits the precession speed to nanoseconds at most for attainable fields. The solution we propose in this article is employing a terahertz magnetic pulse, which we demonstrate is able to modify the dynamics of spin precession within femtoseconds. At the same time, it achieves a redistribution of dipole moment that has the potential to turn dark states bright at a time-scale relevant for the 2DES experiment. \santi{THz electric transients have already been used  in combination with 2DES for the study of Raman spectroscopy, rotational and vibrational spectroscopy of molecules, or phonon excitations \cite{Woerner2013,Nelson2015,Finneran2016,Lu2019}.} By combining such magnetic pulses with a conventional 2DES scheme (see Fig. \ref{Fig1}), we demonstrate that it is possible to study the properties of dark states exhaustively, due to the spectral resolution offered by 2DES.

\section{Model}

To illustrate the mechanism of the magnetic field enhanced 2DES, we use a minimal model containing an electronic ground state $|g,S\rangle$, a spin one singlet electron-hole (e-h) pair $\ket{CT,S} \propto \frac{1}{\sqrt{2}} (\ket{\uparrow}_e \ket{\downarrow}_h -\ket{\downarrow}_e \ket{\uparrow}_h)$ which carries all dipole moment in the system, and triplet states $\ket{CT,T_0} \propto \frac{1}{\sqrt{2}} (\ket{\uparrow}_e \ket{\downarrow}_h +\ket{\downarrow}_e \ket{\uparrow}_h)$, $\ket{CT,T_+} \propto \ket{\uparrow}_e \ket{\uparrow}_h $, $\ket{CT,T_-} \propto \ket{\downarrow}_e \ket{\downarrow}_h $ which are dipole forbidden; i.e. for this model $\hat{\mu} = \mu \ket{CT,S}\bra{g,S} + h.c.$. 

We choose the e-h pair to have low Coulomb binding, \santi{a feature of charge transfer or polaron states in, for example, organic polymer systems \cite{Veldman2009}.} Our choice is based on the importance that these states have in a wide variety of systems ranging from quasi 2D quantum wells to natural and artificial light conducting and harvesting systems \cite{Rao2013a,Gelinas2014a,Chang2015a,Novoderezhkin2007,Ye2014,Snoke2002}. Low Coulomb binding bright states are expected to have low dipole moments. Moreover, owing to the loose nature of the e-h Coulomb binding in charge transfer states, they will be less sensitive to strong magnetic fields. States with just one electron are easier to manipulate with a magnetic field; composite states with stronger binding will typically have a higher dipole moment and will provide a clearer optical signal.

\begin{figure}[h!]
\centering
\includegraphics[width=\columnwidth]{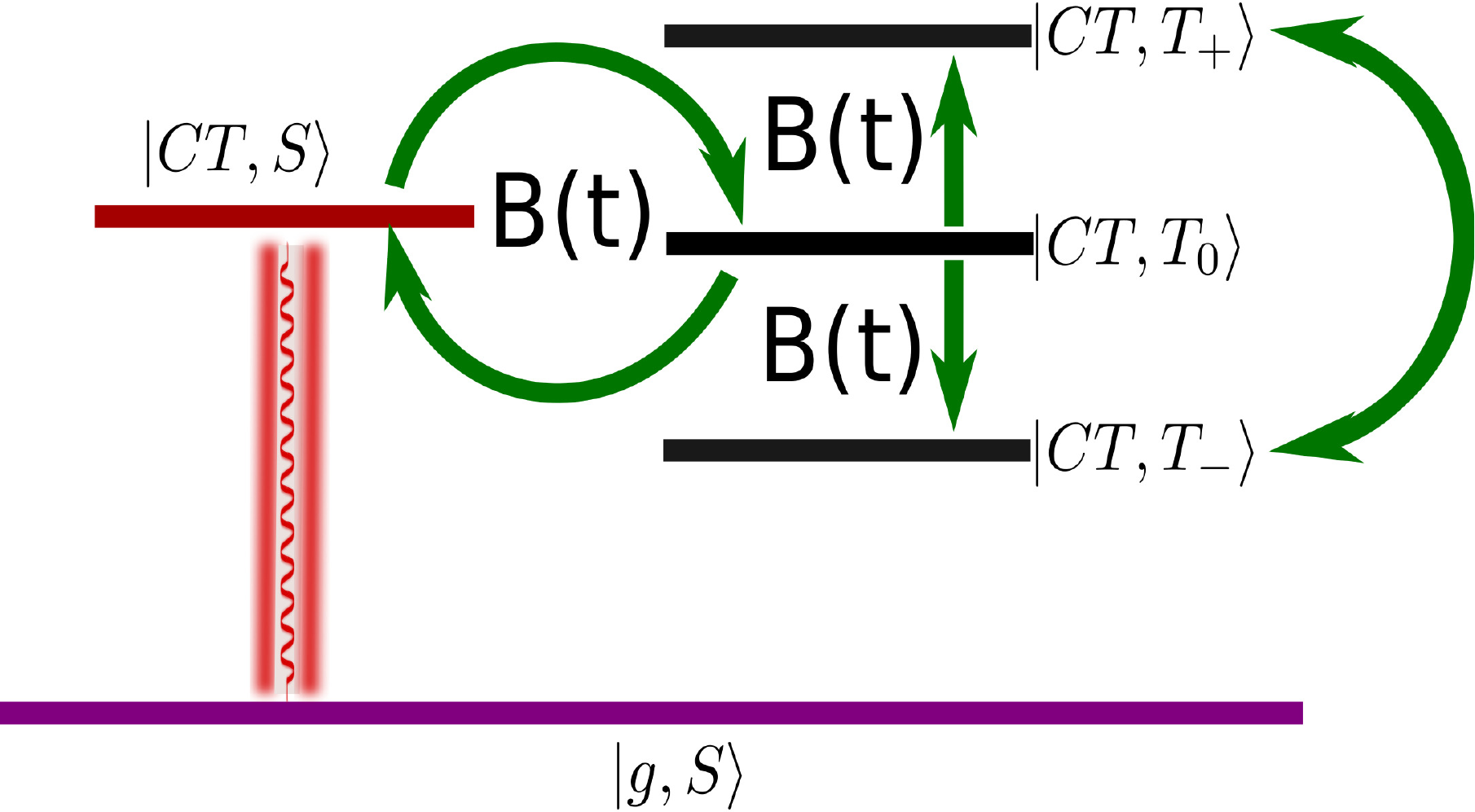}
\caption{Model system employed to demonstrate how a magnetic pulse turns dark states bright in a 2DES experimental scheme. In the model, a singlet charge transfer state is excited through laser light. The three components of the corresponding triplet charge transfer state (degenerated in absence of a magnetic field) are completely dark with respect to the electronic ground state as spin transitions are prohibited upon excitation. An appropriately tuned external magnetic pulse, $B(t)$, is able to induce spin flips on the individual components of an e-h pair by modifying the precession frequency of the spins. Consequently, a magnetic pulse is used to redistribute dipole moment from the singlet state to the spin triplet states. In that way, an initial population in the singlet state can populate first the spin zero triplet state and, subsequently the spin $\pm$ one states, according to Eqs. (\ref{eq2}).}
\label{Fig2}
\end{figure}

The electronic level diagram we employ in our model is depicted in Fig. \ref{Fig2}. At zero magnetic fields the triplet states are degenerate 
\begin{equation}
 H_0=  E_S\ket{CT,S}\bra{CT,S}+ E_T\sum_{i=0,+,-} \ket{CT,T_i}\bra{ CT,T_i}.
\end{equation}
This singlet – triplet structure will be subjected to inter-system crossing (ISC) by the external magnetic pulse, whose dynamics are explained in detail in section \ref{sect3}.

The singlet and triplet charge transfer states are energetically separated ($E_S\neq E_T$) for ultrashort timescales of 2DES experiments, where the decoherence times impose a limit of few picoseconds \cite{Lienau2016}. Natural inter-system crossing can occur as a result of thermal fluctuations of the fine structure, spin-orbit coupling, hyperfine interaction, or wavefunction spatial overlap. These effects appear on much longer (hundred picosecond to microsecond) timescales  
\cite{Kohler2009,Cohen2009,Zhang2012}, or break the degeneracy of the triplet states by a fraction of meV \cite{Cohen2009}. Throughout this paper, only hyperfine interaction will be considered.

In addition to possible coherent singlet-triplet transitions, spin-pair states are subject to the action of an environment composed of electronic and vibrational degrees of freedom causing decoherence and dephasing of the quantum states. This dynamics will be accounted for by Lindblad formalism as described in section \ref{Section4}.

\section{Terahertz inter-system crossing}\label{sect3}

Nowadays, controlling spins with an external magnetic field is technically achievable \cite{Alegre2007,Fuchs2009,London2014}. Terahertz magnetic field pulses have become only recently available \cite{Ferguson2002,Tonouchi2007}, offering the possibility of accessing ultrafast magnetization dynamics on the femtosecond time-scale \cite{Zhao2008,Yamaguchi2010,Nakajima2010,Kampfrath2010,Yamaguchi2013,Kim2014}. Moreover, contrary to what happens with optical radiation, which interacts with valence electrons and interferes with the 2DES target states, terahertz photons have energies around the meV, which will not produce optical transitions \cite{Kampfrath2013}. \santi{The coupling of the electric field associated with the magnetic pulse to the dipoles of the system is expected to be small ($\sim$ meV), consequently it will only induce vibrations on the sample that add-up to the dephasing noise. Furthermore, the small pulse energy, in the order of femtojoules, means that thermal effects are negligible \cite{Yamaguchi2013}. Experiments show that upon interaction with a terahertz pulse, the relatively large electric field associated does not produce appreciable short-term effects, although in the long run, damage is possible \cite{Kampfrath2010,Vicario2013}. This means that for several 2DES with terahertz magnetic pulse cycles the sample is assured to survive. In general, no significant change of the interactions or deformation of the wavefunctions is to be expected even for the fields involved \cite{Scholes2018}. In fact, comparable or stronger electric fields have already been used in combination with 2DES \cite{Loukianov2017}, and applied to organic samples, with positive results. }

To produce spin flips --as it is needed for inter-system crossing-- the coupling induced by the magnetic field has to be strong enough to ensure that the precession of the spins is faster than the 
relaxation time of the states. However, the typical coupling of an electron (hole) to a magnetic field is mediated by the Bohr magneton and the  Land{\'e} factor. The small value of these two factors 
in typical e-h pairs has to be compensated with a magnetic field of at least kiloteslas in order to produce the sub-picosecond inter-system crossing necessary for having an effect on the 2D spectra 
\cite{Oviedo2017}. Such magnetic fields are beyond any experimental instrument available nowadays, and are likely to disrupt the sample. To circumvent the limitation, we propose employing 
terahertz magnetic pulses, whose interaction with the spin components is strong enough to produce the necessary spin flips \cite{Zhao2008,Nakajima2012,Vicario2013,Kim2014}. As we demonstrate in this 
article, the key factor which allows to reduce the magnitude of the fields employed, is not the interaction strength, but rather the frequency with which the magnetic pulse oscillates.

The interaction of a spin with an external oscillating magnetic field with components in the z and x directions ($\vec{B} = B_z \vec{z} + B_x \vec{x}$) is governed by the interaction Hamiltonian for spin pair dynamics \cite{Malla2017}
\begin{equation}
 H = B_e S_e^z + B_h S_h^z + B (S_e^x + S_h^x),
 \label{ham1}
\end{equation}
where each component $B_{e,h}$ references the total magnetic field acting on either the electron or the hole, thus including the hyperfine interaction component felt by each of the spins, which by convention is taken in the z direction. $B_{e,h} (B)$ in Eq. (\ref{ham1}) already includes the coupling of the magnetic field with the spin in the form $g_{i}\mu_B/2$, with g being the Land{\'e} factor and $\mu_B$ the Bohr magneton constant. The Land{\'e} factor is specific for either the electron or the hole. Thus $B_{e} = -\mu_B g_e(B_z + a_eI_e)$, and $B_{h} = -\mu_B g_h(B_z + a_hI_h)$, where I is the hyperfine magnetic field.

The hyperfine component is created by the interaction of the magnetic moment with the atomic nuclei environment surrounding it. At the relevant 2DES time-scale, i.e. femtoseconds, the motion of such nuclei is slow and can be considered static. This fact, together with the large amount of nuclei contributing to the hyperfine field, means that the total hyperfine component can be approximated by a Gaussian distribution \cite{Schulten1978}, where the width reflects the typical strength of the hyperfine component for each material. In the case of most organic materials this width is approximately 1 meV \cite{Flatte2012}. We calculate the hyperfine contribution to the ISC by averaging over 10$^3$ random instances drawn from such Gaussian distribution.
The hyperfine component alone is not enough to produce meaningful inter-system crossing. However, in separating the effect on the electron and the hole, it allows for the external magnetic field to have an amplifying effect which results in greater inter-system crossing \cite{Flatte2016}.

The unitary singlet-triplet dynamics is described by the Schr\"odinger equation for the time evolution of a wavefunction in the singlet-triplet basis spanned by the total spin of the charge transfer state. For reasons of simplicity we omit here possible non-unitary effects, such as dephasing, which will be accounted for in the 2DES simulations. In particular, the dynamics is described by amplitudes S(t), T$_0$(t), T$_+$(t) and T$_-$(t), respectively, i.e. 
\begin{equation}
e^{iE_T t/\hbar} \ket{\psi (t)}= S(t)\ket{CT,S} + \sum_{i=+,-,0}  T_i(t) \ket{CT,T_i}. 
\end{equation}

This base is a logical choice as excitation by incoming pulses in 2DES occurs directly in the singlet state (we assume the impulsive limit in which the laser pulses is much faster than any other timescale in the dynamics). The time evolution of the amplitudes is governed by the equations of motion 
\begin{align}
\begin{split}
{\hbar}\frac{dS(t)}{dt}   &= - i\delta_0(t) T_0(t) - i(E_S-E_T)S(t), \\
{\hbar}\frac{dT_0(t)}{dt} &= -i\delta_0(t) S(t) - i\sqrt{2}B(t)(T_+(t) + T_-(t)), \\
{\hbar}\frac{dT_+(t)}{dt} &= -i {\bar{B}}(t)T_+(t) - i\sqrt{2}B(t)T_0(t), \\ 
{\hbar}\frac{dT_-(t)}{dt} &= i{\bar{B}}(t)T_-(t) - i\sqrt{2}B(t)T_0(t).
\label{eq2}
\end{split}
\end{align}

\begin{figure}
\includegraphics[width=\columnwidth]{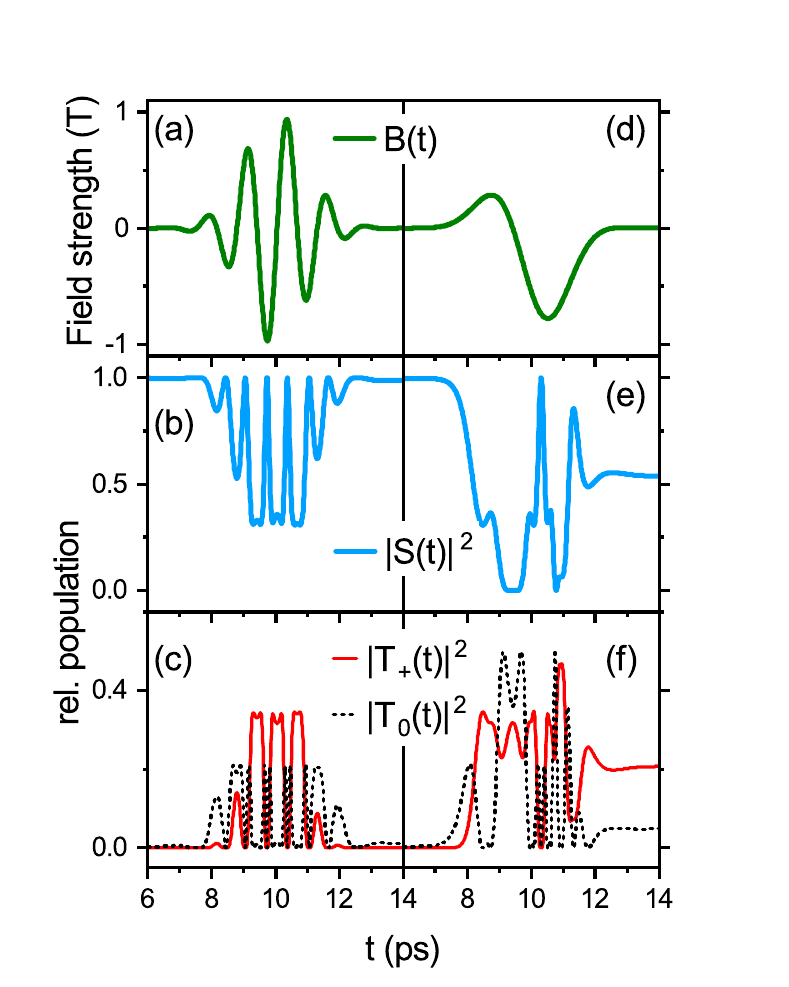}
 \caption{Numerical solution of Eqs. (\ref{eq2}) for a set of singlet-triplet charge transfer states, subject to the effect of a magnetic pulse of the form in Eq. (\ref{eq3}). In the left column, (a) displays a resonant pulse $B \approx \omega$ with $B_0$ = 1 Tesla, which induces coherent Rabi oscillations on an initially populated singlet charge transfer state (b), transferring population to the initially dark triplet states (c). After the pulse ends, no population remains in the triplet state. In the right column, the magnetic pulse is off-resonant ($\omega \approx 0.1B$) with $B_0$ = 1 Tesla, but $\omega = 100$ GHz, as shown in (d). Such a pulse produces coherent population transfer from an initially populated singlet state (e) to initially dark triplet states in (f). In this case, the oscillations not only have a much bigger amplitude than in (a), but also they have more frequencies intermixed. Moreover, triplet states remain partially populated at the end of the pulse. Triplet states $\pm$ 1 have identical population time evolution.}      
 \label{Figmagnetic}
\end{figure}

In Eqs. (\ref{eq2}), $B(t)$ represents the magnetic field, while $\delta_0(t)$ and ${\bar{B}}(t)$ are, respectively, the difference and sum of the interaction of the magnetic field with each of the spins, i.e. 
$\delta_0=\frac{B_e(t) - B_h(t)}{2}$,${\bar{B}}$ =$\frac{B_e(t) + B_h(t)}{2}$. From these equations two conclusions emerge immediately: first, why it is required that the Land{\'e} factor has to be different for 
the electron and the hole, for otherwise $\delta$ = 0 and the dynamics only mix the triplet states. And second: the magnetic field must exhibit z- and x- polarization components. \santi{The reason is that circular polarization allows spin mixing between the singlet and the triplet spin zero components --both polarized along the z-direction-- and also from the triplet spin zero to the triplet $\pm$ one components, which will be Zeeman split resulting in distinct peaks in a 2D map. Here, we consider the possible phase between z- and x- pulse directions to be $\pi/2$, as it will not be important for spin dark states spectroscopy. Nonetheless the effect of finite phase might be worth exploring in a future work, since it can be used to (for example) delay the onset of transfer to certain components of the triplet states, with relevance not only in spectroscopy but also in spintronics.} The previous point relates to the aim of spin state mixing not only between the singlet and triplet spin zero states  --both polarized along the z-direction-- but also among the three triplet spin states, which will be Zeeman split resulting in distinct peaks in a 2D map. \santi{Importantly, it is not necessary for the magnetic pulse to be resonant with the energy difference among singlet and triplet states, though if the condition is met, transfer would happen at lower intensity pulses.}

The magnetic pulse
\begin{align}
\begin{split}
B_x&= B_0 \sin(\omega t-\frac{\pi}{2}) \exp\left[\frac{(t-t_0)^2}{2\sigma_t^2}\right] \\
B_z&= B_0 \sin(\omega t) \exp\left[\frac{(t-t_0)^2}{2\sigma_t^2}\right]
\label{eq3}
\end{split}
\end{align}
has an oscillating component and a Gaussian envelope. The behavior of each spin of the spin-pair in an oscillating magnetic field under strong drive, described by Eqs. (\ref{eq2}) and displayed in Figs. \ref{Figmagnetic}, goes as $\cos(\frac{B_0}{\omega}\cos\omega t)$ for an envelope that is sufficiently wide such that its effect can be neglected for the purposes of dynamics of the spin pair. The ratio $\frac{B_0}{\omega}$ controls the amplitude of the oscillations, while the frequency depends both on the ratio $\frac{B_0}{\omega}$ and on $\omega$. Thus, we speak of two regimes. The first is called resonant and occurs when $B_0 \approx \omega$, while the second is denominated off-resonant \cite{Malla2017}. Notice that here resonance does not have anything to do with energy levels. 

Figs. \ref{Figmagnetic}(a) and (d), show singlet triplet dynamics for two example magnetic pulses. In both cases, the amplitude of the 
field $B_0$ is chosen to be 1 Tesla while its frequency $\omega$ is in an order of magnitude larger in Fig. \ref{Figmagnetic}(a) as compared to Fig. \ref{Figmagnetic}(d).
In both left and right panels on Fig. \ref{Figmagnetic} the initial state witnesses the spin singlet charge transfer state (Figs. \ref{Figmagnetic}(b) and (e)) completely 
populated while the three triplet states have no population at all. Initially, as the pulse interacts with the spin pair, the population is partially transferred to the spin zero triplet charge transfer state and afterwards into the spin $\pm$ 1 triplet charge transfer states, as represented in Figs. \ref{Figmagnetic}(c) and (f) (only the spin +1 case is shown as the spin -1 behaves in exactly the same way). 

From the analysis in Figs. \ref{Figmagnetic} we observe that a resonant pulse produces coherent Rabi oscillations that vanish as the pulse ends. In this case, very clean oscillations in the population of the charge transfer states can be observed, but there will be no final steady state population in the dark triplet charge transfer state (see \ref{Figmagnetic}(c)). However, working out of resonance means that on the one hand, several frequency components show up in the population oscillations and on the other hand that the steady state after the pulse has ended shows population in all four spin charge transfer states, which will decay on a slow timescale. Consequently, effective population transfer to the triplet states occurs with pulses as the one represented in Fig. \ref{Figmagnetic}(f). We conclude that for the purposes of this article where clean oscillations are ideal, resonant pulses are much better suited, while for applications where the interest lays in transferring population and controlling the spin states, off-resonant pulses would be required.

\section{2DES with a magnetic pulse}
\label{Section4}

The development of the theory of nonlinear spectroscopy \cite{Mukamelbook}, together with the progress of ultrafast laser technologies and multidimensional spectroscopic methods \cite{Hamm1998} 
represented a decisive step forward in the study of dynamics in condensed-matter quantum systems. The result was a deeper understanding of processes such as light absorption, energy transport, and 
quantum dynamics in open systems. 2DES represents a recent highlight in this development 
\cite{Jonas2003,Fleming2005,Zigmantas2006,Engel2007,Schlaucohen2011,Plenio2013,Lim2015,Lienau2016,Prior2017,Zigmantas2018}. 

2DES is a powerful technique to study nuclear and electronic correlations between different transitions or initial and final states. It utilizes three ultrashort, spectrally broad laser pulses 
separated by controlled time delays (see Fig. \ref{Fig1}(a)) together with a local oscillator. The Fourier transform of the system response with respect to the coherence time $t_1$ (time between the 
first and second pulses) and with respect to the rephasing time $t_3$ (time between the third pulse and the local oscillator) yields a 2D spectrum in the frequency domain which correlates absorption and 
emission frequencies at each population time $t_2$ (time between the second and third pulses). To increase the number of coherent superpositions between quantum states, broad-band excitation lasers are 
used. Each specific feature in the 2D spectrum then corresponds with one superposition between quantum states and provides real-time information about the both population and coherence 
dynamics in the system.

The simulation of the 2DES-signals requires calculating the third-order non-linear response function $S^{(3)}(t_3,t_2,t_1)$ of the material which relates the driving fields of pulses coming at 
intervals $t_1$, $t_2$ to the induced nonlinear polarization at delay $t_3$ after last pulse. $S^{(3)}(t_3,t_2,t_1)$ is only defined for positive times and reads
\begin{equation}
 \begin{split}
S(t_3,t_2,t_1) &=\left(\frac{i}{{\hbar}}\right)^3 {\rm Tr} \hat{\mu} \mathcal{U}(t_1+t_2+t_3,t_2+t_1)\mu^{(-)}\mathcal{U}(t_1+t_2,t_1) \\& \mu^{(-)} \mathcal{U}(t_1,0)\mu^{(-)} \rho(0)
\label{eq7}
\end{split}
\end{equation}
where $\mu^{(-)} \ldots = [\hat{\mu}, \ldots ] $ is superoperator notation for commutator with dipole $\mu$. And where the $\mathcal{U}(t_a,t_b)$ is evolution superoperator which bring the density 
matrix from time $t_b$ to time $t_a$ between the pulses, $\rho(t_b)= \mathcal{U}(t_a,t_b)\rho(t_b)$, in other words it is the Green function solution of the equation of motion governing the evolution of the density matrix
\begin{equation}
 \frac{d\rho}{dt} =-\frac{i}{{\hbar}}\left[\mathcal{H}(t),\rho\right] + L (\rho),
 \label{eqdensity}
\end{equation}
between two times. Here $L$ describes pure dephasing processes
 \begin{equation}
 L = \sum_{\alpha=S,T} \gamma_{i} \left[ \sigma_{\alpha} \rho(t) \sigma_{\alpha}^\dagger - \frac{1}{2} \left\{ \sigma_{\alpha}^{\dagger}\sigma_{\alpha},\rho(t) \right\}\right],
\label{Eq1}
\end{equation}
Where the fluctuations of singlet state $\sigma_{S}= \ket{CT,S}\bra{CT,S}$ are associated with dephasing rate  $\gamma_{S}$ and (less intense) fluctuations of triplet states
$\sigma_{T}= \sum_{i \in \{0,+,-\}} \ket{CT,T_i}\bra{CT,T_i}$ with dephasing rate  $\gamma_{T}$. This asymmetry originates from the magnetic noise affecting each of the states and to which triplet states can be more resilient. Note however that, since the key point regarding dephasing is that coherence survives long enough for singlet states generation, the main results of this article are unchanged by considering both dephasing rates as equal. The dephasing among triplet states is considered negligible.

The 2D signals are usually displayed in a mixed time-frequency domain 
\begin{equation}
S(\Omega_3, t_2, \Omega_1)=\int_0^{\infty} dt_1  \int_0^{\infty} dt_3 S(t_3,t_2,t_1)e^{i\Omega_3t_3} e^{\pm  i\Omega_1t_1}
\end{equation}
where the + (-) sign  is applied for rephasing (nonrephasing) contributions to signal \cite{Schlaucohen2011,Plenio2013,Lim2015}. The double time dependence $t_a, t_b$ of evolution operator also challenges the proper  definition of absorption spectra. We remain with $t_b=0$ , i.e. we define  $I(\Omega)= \int_0^{\infty} dt e^{i\Omega t}\int_0^{\infty} {\rm Tr} \hat{\mu} \mathcal{U}(t,0)\mu^{(-)} \rho(0)$ when the absorption spectrum is simply obtained from 2D by integrating over $\Omega_3$, i.e. by using projection-slice theorem 
$\int_0^{\infty} d\Omega_3  S(\Omega_3, t_2=0, \Omega_1)=S(t_3=0, t_2=0, \Omega_1)\propto I(\Omega_1)$.

For the coherent evolution $\mathcal{H}(t)$ we propose a scheme that includes a magnetic pulse that will be acting on the sample through the duration of the laser pulses sequence (see 
Fig.~\ref{Fig1}); consequently, the Hamiltonian depends explicitly on time (and evolution operator $\mathcal{U}$ indeed depends on both initial and final time). Concretely, the Hamiltonian for the system interacting with the magnetic pulse reads
\begin{equation}
\begin{split}
 \mathcal{H}(t) &= E_S\ket{S}\bra{S} + E_{T}\ket{T_0}\bra{T_0}  + (E_{T} + Z(t))\ket{T_+}\bra{T_+} \\ &  + (E_{T} - Z(t) )\ket{T_-}\bra{T_-}+ \mathcal{H_I^B(t)},
\end{split}
\end{equation}
where $Z(t)={\bar{B}}$ describes the (time-dependent) Zeeman splitting while $\mathcal{H_I^B}(t)$ describes part of the interaction which induces coherent transport among energy levels, and which reads
\begin{equation}
 \mathcal{H_I^B}(t) = \delta_0 \ket{S}\bra{T_0} + B \ket{T_0}\left( \bra{T_+} + \bra{T_-} \right) + c.c..
\end{equation}

Eq. \ref{eqdensity} describes the evolution of the density matrix interacting with impulsive laser pulses \cite{Plenio2013}. Note that while the laser pulses are treated perturbatively, the magnetic pulse interaction with the system is treated exactly. The evolution is averaged over the phase of magnetic fields (as these are not phase synchronized with laser fields).

\section{Spin dark states detection}

We simulate the dynamics of a singlet-triplet system interacting with a resonant magnetic pulse in different spectroscopic configurations. \santi{We choose Landé factor values that mimic organic compounds used for solar cells fabrication, with $g_e = g_h \approx$ 2, and $\Delta g = 10^{-3}$ when the difference is relevant for dynamics}. The magnetic pulse is chosen to have amplitude $B_0$ = 1 T, 
oscillation frequency $\omega$ = 1 THz, and a Gaussian envelope that guarantees the pulse has approximately constant amplitude through the duration of the laser sequence simulation, namely $>$ 400 fs. 
Initially, dipole moment is associated only with the singlet state, meaning that the triplet state is completely dark in all the simulations. Consequently, in the absence of a magnetic field, the only 
contribution comes from the singlet charge transfer state absorption. We note that the model presented here is not substantially altered by marginal singlet-triplet coupling or non-zero dipole moment 
of the dark state.

\begin{figure}
\includegraphics[width=10cm]{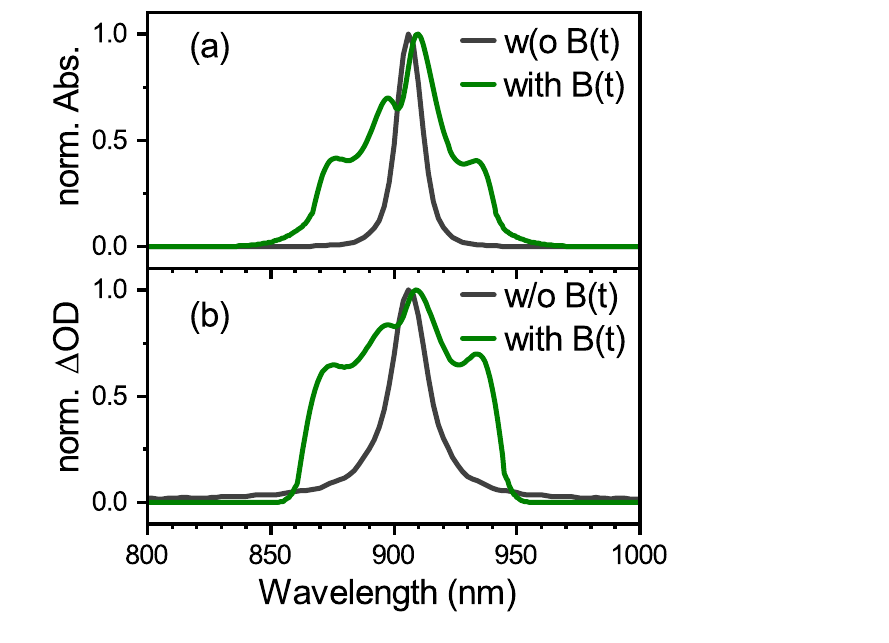}
\caption{Numerical simulation of absorption spectra (a) and transient absorption spectra (b) in the model with a singlet and triplet charge transfer states with energies in 885.6 (1.4) and 892 (1.39) 
nm (eV) respectively, in the absence (grey) and presence (green) of a magnetic pulse of frequency 1 THz and amplitude 1 Tesla, with an envelope that guarantees the pulse duration for the whole 
experiment. Both plots are calculated at 400 fs population time. Notice that the peaks have been scaled to appear with similar height. In the absence of magnetic interaction only an absorption peak for the singlet charge transfer state shows in the 
spectrum while in the presence of interaction a complicated peaks structure develops.}
\label{Figure2D2}
\end{figure}

2DES is a costly technique and its use has to be well-motivated when simpler spectroscopy techniques might suffice. In Fig. \ref{Figure2D2}(a), we plot the absorption spectrum in the absence and in 
the presence of a magnetic pulse. These results demonstrate that in absence of B-induced interaction, there is only one absorption peak corresponding with the singlet state. Yet in the presence of 
such an interaction, a complicated spectral signature emerges (see green line in Fig. \ref{Figure2D2}(a)). \santi{Though a four peak structure emerges, as expected from the singlet-triplet peak splitting due to the magnetic field}, any further analysis is hampered by the convoluted character of the signal, leading to complicated lineshapes. \santi{Note that the slight displacement of the central (main) peak is due to the interaction of the singlet state with the magnetic pulse.}

Moving on to time-resolved methods, transient absorption spectroscopy (pump-probe) yields information about the energy of the quantum states and the transition and relaxation rates among them. In Fig. 
\ref{Figure2D2}(b), we plot the pump-probe spectrum at a pump-probe delay of 400 fs. In the absence of a magnetic field, the lineshape of the singlet absorption peak is a well-defined Lorentzian. Similar to the case of the linear 
spectrum discussed above, the spectral features are convoluted in a non-trivial manner in the presence of a magnetic field. \santi{The presence of the same four peaks as in absorption spectra Fig. \ref{Figure2D2} but with different relative intensity indicates energy transfer and, therefore, electronic coupling.} Since the only source of peak splitting is the magnetic pulse, resolution 
depends on the interaction strength of the magnetic pulse with the electron and the hole, which is in principle unknown. The information about the interaction is encoded in the dipole moment redistribution and the coherence transfer among states, as a consequence of the magnetic pulse, and conspicuous either as oscillations in the population peaks or as distinctive in 2DES.

\begin{figure}
\includegraphics[width=\columnwidth]{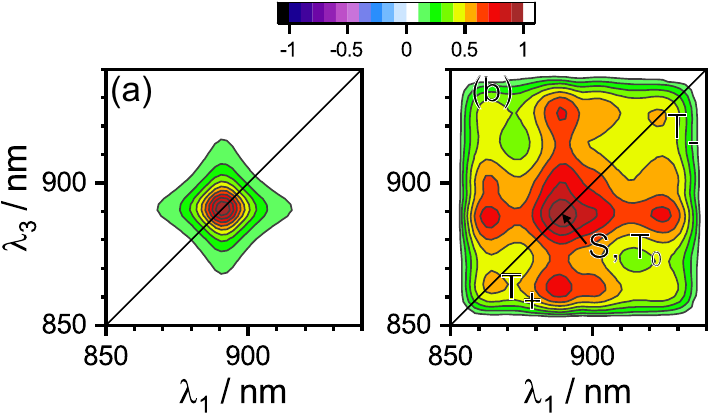}
\caption{Numerical simulation of a 2D electronic spectrum according to the model with a singlet and triplet charge transfer states with energies in 885.6 (1.4) and 892 (1.39) nm (eV) respectively. In (a), in the 
absence of a magnetic pulse the absorption peak corresponding to the singlet charge transfer state. In (b), the action of a magnetic pulse of frequency 1 THz and amplitude 1 Tesla, with an envelope 
that guarantees the pulse duration for the whole experiment, permits dipole moment redistribution from the singlet to the various triplet charge transfer states, as well as ground-excited states coherences transfer, as evidenced by several diagonal peaks and non-diagonal structures which demonstrate the presence of coherent transport among different quantum states. The latter are Zeeman split, with a gap of 1.1 (1) nm (meV). The coupling strength of the magnetic pulse to the charge transfer states is 112 (100) nm (meV) while the lifetimes are 80 fs for the singlet charge transfer state and 200 fs for the triplet components. Both plots are at population time 400 fs.}
\label{Figure2D1}
\end{figure}

In Fig. \ref{Figure2D1}, we plot the real part of the spectrum resulting from the sum of all rephasing and non-rephasing components of ground state bleaching (GSB) and stimulated emission (SE). The 
polarization of the excitation pulses was set to all-parallel. Such a configuration yields the strongest overall 2D-signal, and also contains coherence dynamics from intramolecular
states \cite{Mancal2012}, 
which is readily accounted for in simulations. Fig. \ref{Figure2D1}(a) displays a characteristic star-shaped 2D-peak corresponding with the singlet state which in absence of a magnetic pulse is the only 
bright state of the system. Interaction of the states with a magnetic pulse through the duration of the 2DES sequence produces a spectrum richly populated with distinctive features in Fig. 
\ref{Figure2D1}(b). 

Focusing the analysis on the central part of the spectrum, we observe several peaks along the diagonal (see also Figs. \ref{Figure2D2}(a) and(b)) that correspond to absorption peaks from 
different states composing the system. We are witnessing the triplet dark states mixed with the singlet state by the magnetic pulse. In addition, the numerous crosspeaks correspond to the different 
coherences that gather the information of the corresponding redistribution of dipole moment among states and the coherence transfer between singlet and triplet ground state to excited state coherences.

We have demonstrated that 2DES can reveal the presence and dynamics of dark states if supplemented with a magnetic pulse. As a next step, we reconstruct the properties that characterize these dark states. In Eq. 
(\ref{eq6}), we present the original Hamiltonian which needs to be reconstructed. The diagonal terms are the energies of the states prior to the interaction with the magnetic pulse with an extra term 
we name Z accounting for the Zeeman splitting. The non-diagonal terms are the different interactions according to Eqs. (\ref{eq2}). Conversely, the  diagonal peaks in the 2D spectrum in Fig. 
\ref{Figure2D1}(b) provide information about the exciton energies, namely, about the eigenvalues of Eq. (\ref{eq6}). Therefore, to reconstruct the original energies we need the interaction terms of 
Eq. (\ref{eq6}) and then revert the diagonalization procedure.

\begin{equation}
 \mathcal{H} = \left(
\begin{array}{cccc}
S & 0 & A & 0 \\
 0 &  T_{-} - Z & B & 0 \\
 A & B & T_{0} & B \\
0 & 0 & B & T_{+} + Z
\end{array}
\right).
\label{eq6}
\end{equation}

The diagonal peaks in Fig. \ref{Figure2D1}(b) are located at 866 (1.432), 882 (1.406), 895 (1.385), and 920 (1.348) nm (eV). The information about the interaction terms is encoded in the 
non-diagonal peaks of Fig. \ref{Figure2D1}(b). The location of these peaks tells us about between which two states the coherence transport is happening. The frequency of oscillation of these peaks in 
population time encodes the information about the interaction strength between energy levels. Hence, after multi-exponential fitting to get rid of dephasing, the Fourier transform in population time 
of the non-diagonal peaks 
provides us with the interaction terms. These give us the following numbers: 11 (0.009) nm (eV) for A and 16 (0.013) for B. With them we obtain that the interaction of the magnetic 
pulse with the electron and the hole is 155 (0.124) nm (eV), which gives us a Zeeman splitting of 8 (0.006) nm (meV) and a reconstructed states at 880 (1.41), 894 (1.387), 894.6 (1.386), and 895 
(1.385) nm (eV). Hence we are able to reconstruct the original states 
with an error smaller than 2\%. 

Note that the information required to perform the full reconstruction of the system’s Hamiltonian is attainable neither from linear spectroscopy nor from transient absorption spectroscopy, \santi{which can only reveal the presence of additional states with energy transfer among them}. 2DES on the other hand, fully resolves the peak structure and is necessary to determine the properties of the sates.

\section{Conclusions}

Summarizing, we propose and test numerically a modified version of 2DES in which a terahertz magnetic pulse is employed to create coherent population transfer from a bright spin singlet electron-hole 
pair to the components of the corresponding triplet state. The magnetic pulse is able to modify the spins precession in a time-scale relevant for 2DES. The effect can be understood as a dipole moment 
redistribution with coherence transfer that allows transitions from the ground state to the triplet states, which then appear as distinctive peaks in the 2D-spectrum. State reconstruction from the position of the peaks and 
the oscillation frequencies of the coherences allows one to infer the properties of the original states, a feature that simpler spectroscopic techniques do not allow either for lack of information or 
lack of resolution.

The magnetic pulse employed to demonstrate the feasibility of the proposal is realistic by today standards. The parameters describing the charge transfer states are as well within the typical range. 
Tuning the magnetic pulse so as to produce the maximum effect requires some work, as it is the balance between the amplitude of the pulse and the frequency that dictates the transfer rate among states 
and the frequency at which the coherences will oscillate. Nonetheless, the split of peaks and the unveiling of spin dark states is observed for almost any sensible magnetic pulse; it is therefore easy to 
obtain a first estimation that permits fine-tuning the experiment. The strength with which the magnetic pulse couples to the spin states depends on the nature of the later, and has to be elucidated 
experimentally. However, knowing the value of this coupling is necessary in order to reconstruct the in principle unknown energies of the states analyzed. Hence the need for 2DES experiments is 
warranted.

Notice that different kinds of dark states with different spin configurations will have different equations of motion than Eq. (\ref{eq2}), and the constraints imposed by the Land{\'e} factor will not 
necessarily apply. However, the background physics remains unchanged, and any spin dark state can be manipulated so, being it singlet or triplet. Therefore, the conclusions obtained in this article are valid for a wide class of 
dark states.

\section{Acknowledgement}
S.O.C. and J.P. are grateful for financial support from MCIU 
(SPAIN), including FEDER funds: FIS2015-69512-R and PGC2018-097328-B-100 together with Fundaci{\'o}n S{\'e}neca (Murcia, Spain) Project No. 19882/GERM/15. S.O.C.is supported by
the Fundación Ramón Areces postdoctoral fellowship (XXXI edition of grants for Postgraduate Studies in Life and Matter
Sciences in Foreign Universities and Research Centers 2019/2020). F.\v{S}. acknowledges support by Czech Science Foundation
(Grant No. 17-22160S). F.\v{S}. and J.H. acknowledge the mobility project “Exciton-exciton annihilation probed by non-linear spectroscopy ” (MSMT Grant No. 8J19DE009, DAAD-Projekt 57444962). J.H. acknowledges funding by the Deutsche Forschungsgemeinschaft (DFG, German Research Foundation) under Germany's Excellence Strategy EXC-2089. 
%

\end{document}